\documentclass[12pt,titlepage]{article}
\usepackage{amsfonts}
\usepackage{amsmath}
\usepackage{amssymb}
\usepackage[dvips]{graphicx}

\begin{document}

\title{On the nonminimal vector coupling in the Duffin-Kemmer-Petiau theory
and the confinement of massive bosons by a linear potential}
\author{T.R. Cardoso\thanks{%
cardoso@feg.unesp.br }, L.B. Castro\thanks{%
benito@feg.unesp.br }, A.S. de Castro\thanks{%
castro@pq.cnpq.br.} \\
\\
UNESP - Campus de Guaratinguet\'{a}\\
Departamento de F\'{\i}sica e Qu\'{\i}mica\\
12516-410 Guaratinguet\'{a} SP - Brazil}
\date{}
\maketitle

\begin{abstract}
Vector couplings in the Duffin-Kemmer-Petiau theory are revised. It is shown
that minimal and nonminimal vector potentials behave differently under
charge-conjugation and time-reversal transformations. In particular, it is
shown that nonminimal vector potentials have been erroneously applied to the
description of elastic meson-nucleus scatterings and that the space
component of the nonminimal vector potential plays a crucial role for the
confinement of bosons. The DKP equation with nonminimal vector linear
potentials is mapped into the nonrelativistic harmonic oscillator problem
and the behavior of the solutions for this sort of DKP oscillator is
discussed in detail. Furthermore, the absence of Klein's paradox and the
localization of bosons in the presence of nonminimal vector interactions are
discussed.
\end{abstract}

\section{Introduction}

The first-order Duffin-Kemmer-Petiau (DKP) formalism \cite{pet}-\cite{kem}
describes spin-0 and spin-1 particles and has been used to analyze
relativistic interactions of spin-0 and spin-1 hadrons with nuclei as an
alternative to their conventional second-order Klein-Gordon and Proca
counterparts. The onus of equivalence between the formalisms represented an
objection to the DKP theory for a long time and only recently it was shown
that they yield the same results in the case of minimally coupled vector
interactions, on the condition that one correctly interprets the components
of the DKP spinor \cite{now}-\cite{lun}. However, the equivalence between
the DKP and the Proca formalisms has already a precedent \cite{mr}. The
equivalence does not maintain if one considers partially conserved currents
\cite{FIS} and the DKP formalism proved to be better than the Klein-Gordon
formalism in the analysis of the $K_{l3}$ decays, the decay-rate ratio $%
\Gamma (\eta \rightarrow \gamma \gamma )/\Gamma (\pi ^{0}\rightarrow \gamma
\gamma )$, and level shifts and widths in pionic atoms \cite{fis1}.
Furthermore, the DKP formalism enjoys a richness of couplings not capable of
being expressed in the Klein-Gordon and Proca theories. A number of
different couplings in the DKP formalism, with scalar and vector couplings
in analogy with the Dirac phenomenology for proton-nucleus scattering, has
been employed in the phenomenological treatment of the elastic meson-nucleus
scattering at medium energies with a better agreement to the experimental
data when compared to the Klein-Gordon and Proca based formalisms \cite{cla1}%
-\cite{cla2}. On the other hand, the DKP\ theory has also experienced a
renewed interest due to the discovery of a new conserved vector current \cite%
{now}, \cite{gho1}-\cite{gho6}, whose positive-definite time component would
be a candidate to a probability current, and as a bonus a hope for avoiding
Klein's paradox for bosons \cite{gho6}. However, it has been shown that the
proposed new current is a fiasco as a probability current \cite{stru}. An
effort to disembarrass the status of that new current was done \cite{dat}
but in \cite{tati} it was shown to be indefensible. In Ref. \cite{tati} it
also was shown that Klein's s paradox may exist in the DKP theory with
minimally coupled vector interactions. The DKP theory has also experienced a
renewal of life in the context of applications to quantum chromodynamics
\cite{gri}, covariant Hamiltonian dynamics \cite{kan}, relativistic phase
space \cite{fer}, curved space-time \cite{red}, causal approach \cite{lun},
\cite{LUN}, superluminal tunneling \cite{gho4}, Bohm model \cite{gho2}, \cite%
{gho5}, \cite{stru}, tunneling time \cite{bon}, S-matrix \cite{pim},
five-dimensional Galilean invariance \cite{mon}, pseudoclassical mechanics
\cite{CAS1}, Bose-Einstein condensation \cite{CAS2}, homogeneous magnetic
field \cite{dai}, Aharonov-Casher phase \cite{swa}, Aharonov-Bohm potential
\cite{BOU1}, position-dependent mass and vector step potential \cite{mer2},
time-dependent mass and time-dependent vector fields \cite{mer1}, tensor DKP
oscillator (tensor coupling with a linear potential) \cite{deb}-\cite{kas}
and its thermodynamics properties \cite{BOU4}, vector DKP oscillator
(nonminimal vector coupling with a quadratic potential \cite{Ait} and
minimal plus nonminimal vector couplings with a linear potential \cite{kuli}%
) sextic oscillator (tensor coupling with a linear plus a cubic potential)
\cite{yas1}, vector step potential \cite{gho6}, \cite{CHE}, \cite{tati},
vector Woods-Saxon potential \cite{bou}, vector deformed Hulthen potential
\cite{yas}, vector square well \cite{ned2}, vector Coulomb potentials \cite%
{BOZ10}, \cite{kas},\cite{ned2}-\cite{gon} and nonminimal vector step
potentials \cite{ccc2}.

The main purpose of the present article is to report on the properties of
the DKP theory with the nonminimal vector coupling interaction. Nonminimal
vector potentials, added by other kinds of Lorentz structures, have already
been used successfully in a phenomenological context for describing the
scattering of mesons by nuclei \cite{cla1}-\cite{kal}, \cite{kur1}, \cite%
{cla2}. In this paper it is shown that charge-conjugation and time-reversal
symmetries have some special features not displayed by minimal vector
potentials, in particular nonminimal vector potentials do not couple to the
charge. It is also shown that nonminimal vector couplings have been used
improperly in the phenomenological description of elastic meson-nucleus
scatterings \cite{cla1}-\cite{kal}, \cite{kur1}, \cite{cla2}. Furthermore,
nonminimal vector potentials can be used as a model for confining bosons and
that linear potentials lead to a sort of relativistic DKP oscillator. This
article is organized as follows. In Sec. II we present the general DKP
equation, discuss conditions on the interactions which lead to a conserved
current and effects of parity, charge-conjugation and time-reversal
transformations on the vector Lorentz structures. Adopting a specific
representation for the DKP matrices, we set up the one-dimensional equations
for the components of the DKP spinor (IIA for the spin-0 sector and IIB for
the spin-1 sector) in the presence of minimal and nonminimal vector
interactions. \ We point out that the space component of the nonminimal
vector potential can not be absorbed into the spinor, as diffused in the
literature. Beyond that, we show that the space component of the nonminimal
vector potential could be irrelevant for the formation of bound states for
potentials vanishing at infinity but its presence is a sine qua non
condition for confinement. In Sec. III we specialize to nonminimal vector
linear potentials and discuss the solutions of the vector DKP oscillator in
detail. The relevance of the nonminimal vector potential for the confinement
of bosons is reinforced. An apparent paradox related to the localization of
bosons in the presence of strong potentials is solved by introducing the
concepts of effective mass and effective Compton wavelength. Finally, in
Sec. IV we draw conclusions.

\section{The DKP equation and the vector couplings}

The DKP equation for a free boson is given by \cite{kem} (with units in
which $\hbar =c=1$)%
\begin{equation}
\left( i\beta ^{\mu }\partial _{\mu }-m\right) \psi =0  \label{dkp}
\end{equation}%
\noindent where the matrices $\beta ^{\mu }$\ satisfy the algebra%
\begin{equation}
\beta ^{\mu }\beta ^{\nu }\beta ^{\lambda }+\beta ^{\lambda }\beta ^{\nu
}\beta ^{\mu }=g^{\mu \nu }\beta ^{\lambda }+g^{\lambda \nu }\beta ^{\mu }
\label{beta}
\end{equation}%
\noindent and the metric tensor is $g^{\mu \nu }=\,$diag$\,(1,-1,-1,-1)$.
The algebra expressed by (\ref{beta}) generates a set of 126 independent
matrices whose irreducible representations are a trivial representation, a
five-dimensional representation describing the spin-0 particles and a
ten-dimensional representation associated to spin-1 particles. The DKP
spinor has an excess of components and the theory has to be supplemented by
an equation which allows to eliminate the redundant components. That
constraint equation is obtained by multiplying the DKP equation by $1-\beta
^{0}\beta ^{0}$, namely%
\begin{equation}
i\beta ^{j}\beta ^{0}\beta ^{0}\partial _{j}\psi =m\left( 1-\beta ^{0}\beta
^{0}\right) \psi ,\quad j\text{ \ runs from 1 to 3}  \label{vin1}
\end{equation}%
This constraint equation expresses three (four) components of the spinor by
the other two (six) components and their space derivatives in the scalar
(vector) sector so that the superfluous components disappear and there only
remain the physical components of the DKP theory. The second-order
Klein-Gordon and Proca equations are obtained when one selects the spin-0
and spin-1 sectors of the DKP theory.

A well-known conserved four-current is given by
\begin{equation}
J^{\mu }=\frac{1}{2}\bar{\psi}\beta ^{\mu }\psi  \label{corrente}
\end{equation}%
\noindent where the adjoint spinor $\bar{\psi}$ is given by%
\begin{equation}
\bar{\psi}=\psi ^{\dagger }\eta ^{0}  \label{adj}
\end{equation}%
with%
\begin{equation}
\eta ^{\mu }=2\beta ^{\mu }\beta ^{\mu }-g^{\mu \mu }\;\text{\textrm{(no
summation)}}  \label{etamu}
\end{equation}%
in such a way that $\left( \eta ^{0}\beta ^{\mu }\right) ^{\dagger }=\eta
^{0}\beta ^{\mu }$ (the matrices $\beta ^{\mu }$ are Hermitian with respect
to $\eta ^{0}$). The time component of this current is not positive definite
but it may be interpreted as a charge density. The factor 1/2 multiplying $%
\bar{\psi}\beta ^{\mu }\psi $, of no importance regarding the conservation
law, is in order to hand over a charge density conformable to that one used
in the Klein-Gordon theory and its nonrelativistic limit (see e.g. \cite{puk}%
). Then the normalization condition $\int d\tau \,J^{0}=\pm 1$ can be
expressed as%
\begin{equation}
\int d\tau \,\bar{\psi}\beta ^{0}\psi =\pm 2  \label{norm}
\end{equation}%
where the plus (minus) sign must be used for a positive (negative) charge,
and the expectation value of any observable $\mathcal{O}$ may be given by
\begin{equation}
\left\langle \mathcal{O}\right\rangle =\frac{\int d\tau \,\bar{\psi}\beta
^{0}\mathcal{O}\psi }{\int d\tau \,\bar{\psi}\beta ^{0}\psi }  \label{exp}
\end{equation}%
where $\mathcal{O\ }$ must be Hermitian with respect to $\beta ^{0}$, namely
$\left( \beta ^{0}\mathcal{O}\right) ^{\dagger }=\beta ^{0}\mathcal{O}$, for
insuring that $\left\langle \mathcal{O}\right\rangle $ is a real quantity.

With the introduction of interactions, the DKP equation can be written as%
\begin{equation}
\left( i\beta ^{\mu }\partial _{\mu }-m-V\right) \psi =0  \label{dkp2}
\end{equation}%
where the more general potential matrix $V$ is written in terms of 25 (100)
linearly independent matrices pertinent to the five(ten)-dimensional
irreducible representation associated to the scalar (vector) sector. In the
presence of interactions $J^{\mu }$ satisfies the equation%
\begin{equation}
\partial _{\mu }J^{\mu }+\frac{i}{2}\bar{\psi}\left( V-\eta ^{0}V^{\dagger
}\eta ^{0}\right) \psi =0  \label{corrent2}
\end{equation}%
Thus, if $V$ is Hermitian with respect to $\eta ^{0}$ then the four-current
will be conserved. The potential matrix $V$ can be written in terms of
well-defined Lorentz structures. For the spin-0 sector there are two scalar,
two vector and two tensor terms \cite{gue}, whereas for the spin-1 sector
there are two scalar, two vector, a pseudoscalar, two pseudovector and eight
tensor terms \cite{vij}. The tensor terms have been avoided in applications
because they furnish noncausal effects \cite{gue}-\cite{vij}. Considering
only the vector terms, $V$ is in the form%
\begin{equation}
V=\beta ^{\mu }A_{\mu }^{\left( 1\right) }+i[P,\beta ^{\mu }]A_{\mu
}^{\left( 2\right) }  \label{pot}
\end{equation}%
where $P$ is a projection operator ($P^{2}=P$ and $P^{\dagger }=P$) in such
a way that $\bar{\psi}P\psi $ behaves as a scalar and $\bar{\psi}[P,\beta
^{\mu }]\psi $ behaves like a vector. Notice that the vector potential $%
A_{\mu }^{\left( 1\right) }$ is minimally coupled but not $A_{\mu }^{\left(
2\right) }$. One very important point to note is that this matrix potential
leads to a conserved four-current but the same does not happen if instead of
$i[P,\beta ^{\mu }]$ one uses either $P\beta ^{\mu }$ or $\beta ^{\mu }P$,
as in \cite{cla1}-\cite{kal}, \cite{kur1}, \cite{cla2}, \cite{Ait}). As a
matter of fact, in Ref. \cite{cla1} is mentioned that $P\beta ^{\mu }$ and $%
\beta ^{\mu }P$ produce identical results.

If the terms in the potential $V$ are time-independent one can write $\psi (%
\vec{r},t)=\phi (\vec{r})\exp (-iEt)$, where $E$ is the energy of the boson,
in such a way that the time-independent DKP equation becomes%
\begin{equation}
\left[ \beta ^{0}\left( E-A_{0}^{\left( 1\right) }\right) +i\beta ^{i}\left(
\partial _{i}+iA_{i}^{\left( 1\right) }\right) -\left( m+i[P,\beta ^{\mu
}]A_{\mu }^{\left( 2\right) }\right) \right] \phi =0  \label{DKP10}
\end{equation}%
In this case \ $J^{\mu }=\bar{\phi}\beta ^{\mu }\phi /2$ does not depend on
time, so that the spinor $\phi $ describes a stationary state. Note that the
time-independent DKP equation is invariant under a simultaneous shift of $E$
and $A_{0}^{\left( 1\right) }$, such as in the Schr\"{o}dinger equation, but
the invariance does not maintain regarding $E$ and $A_{0}^{\left( 2\right) }$%
. Eq. (\ref{DKP10}) for the characteristic pair $(E_{k},\phi _{k})$ can be
written as%
\begin{equation}
\left( E_{k}-A_{0}^{\left( 1\right) }\right) \beta ^{0}\phi _{k}+i\left(
\overrightarrow{\partial _{i}}+iA_{i}^{\left( 1\right) }\right) \beta
^{i}\phi _{k}-m\phi _{k}-iA_{\mu }^{\left( 2\right) }[P,\beta ^{\mu }]\phi
_{k}=0  \label{orto1}
\end{equation}%
and its adjoint form, by changing $k$ by $k^{\prime }$, as%
\begin{equation}
\left( E_{k^{\prime }}-A_{0}^{\left( 1\right) }\right) \bar{\phi}_{k^{\prime
}}\beta ^{0}-i\bar{\phi}_{k^{\prime }}\beta ^{i}\eta ^{0}\left(
\overleftarrow{\partial _{i}}-iA_{i}^{\left( 1\right) }\right) -m\bar{\phi}%
_{k^{\prime }}\eta ^{0}+iA_{\mu }^{\left( 2\right) }\bar{\phi}_{k^{\prime
}}\eta ^{0}[P,\beta ^{\mu }]^{\dagger }=0  \label{orto2}
\end{equation}%
By multiplying (\ref{orto1}) from the left by $\bar{\phi}_{k^{\prime }}$ and
(\ref{orto2}) from the right by $\eta ^{0}\phi _{k}$ leads to%
\begin{equation}
\left( E_{k}-A_{0}^{\left( 1\right) }\right) \bar{\phi}_{k^{\prime }}\beta
^{0}\phi _{k}+i\bar{\phi}_{k^{\prime }}\beta ^{i}\left( \overrightarrow{%
\partial _{i}}+iA_{i}^{\left( 1\right) }\right) \phi _{k}-m\bar{\phi}%
_{k^{\prime }}\phi _{k}-iA_{\mu }^{\left( 2\right) }\bar{\phi}_{k^{\prime
}}[P,\beta ^{\mu }]\phi _{k}=0  \label{orto3}
\end{equation}%
and%
\begin{equation}
\left( E_{k^{\prime }}-A_{0}^{\left( 1\right) }\right) \bar{\phi}_{k^{\prime
}}\beta ^{0}\phi _{k}-i\bar{\phi}_{k^{\prime }}\beta ^{i}\left(
\overleftarrow{\partial _{i}}-iA_{i}^{\left( 1\right) }\right) \phi _{k}-m%
\bar{\phi}_{k^{\prime }}\phi _{k}+iA_{\mu }^{\left( 2\right) }\bar{\phi}%
_{k^{\prime }}\eta ^{0}[P,\beta ^{\mu }]^{\dagger }\eta ^{0}\phi _{k}=0
\label{orto4}
\end{equation}%
respectively. Subtracting (\ref{orto4}) from (\ref{orto3}) and considering
that the spinors fit boundary conditions such that%
\begin{equation}
\int d\tau \,\partial _{i}\left( \bar{\phi}_{k}\beta ^{i}\phi _{k^{\prime
}}\right) =0  \label{orto5}
\end{equation}%
one gets%
\begin{equation}
\left( E_{k}-E_{k^{\prime }}\right) \int d\tau \,\bar{\phi}_{k}\beta
^{0}\phi _{k^{\prime }}=0  \label{orto6}
\end{equation}%
Eq. (\ref{orto6}) is an orthogonality statement applying to the DKP
equation. Any two stationary states with distinct energies and subject to
suitable boundary conditions are orthogonal in the sense that
\begin{equation}
\int d\tau \,\bar{\phi}_{k}\beta ^{0}\phi _{k^{\prime }}=0,\quad \mathrm{for}%
\quad E_{k}\neq E_{k^{\prime }}  \label{orto7}
\end{equation}%
In addition, in view of (\ref{norm}) the spinors $\phi _{k}$ and $\phi
_{k^{\prime }}$ are said to be orthonormal if%
\begin{equation}
\int d\tau \,\bar{\phi}_{k}\beta ^{0}\phi _{k^{\prime }}=\pm 2\delta
_{E_{k}E_{k^{\prime }}}  \label{orto8}
\end{equation}

The DKP equation is invariant under the parity operation, i.e. when $\vec{r}%
\rightarrow -\vec{r}$, if $A_{i}^{\left( 1\right) }$ and $A_{i}^{\left(
2\right) }$ change sign, whereas $A_{0}^{\left( 1\right) }$ and $%
A_{0}^{\left( 2\right) }$ remain the same. This is because the parity
operator is $\mathcal{P}=\exp (i\delta _{P})P_{0}\eta ^{0}$, where $\delta
_{P}$ is a constant phase and $P_{0}$ changes $\vec{r}$ into $-\vec{r}$.
Because this unitary operator anticommutes with $\beta ^{i}$ and $[P,\beta
^{i}]$, they change sign under a parity transformation, whereas $\beta ^{0}$
and $[P,\beta ^{0}]$, which commute with $\eta ^{0}$, remain the same. Since
$\delta _{P}=0$ or $\delta _{P}=\pi $, the spinor components have definite
parities. The charge-conjugation operation changes the sign of the minimal
interaction potential,\textit{\ i.e. }changes the sign of \ $A_{\mu
}^{\left( 1\right) }$. This can be accomplished by the transformation $\psi
\rightarrow \psi _{c}=\mathcal{C}\psi =CK\psi $, where $K$ denotes the
complex conjugation and $C$ is a unitary matrix such that $C\beta ^{\mu
}=-\beta ^{\mu }C$. The matrix that satisfies this relation is $C=\exp
\left( i\delta _{C}\right) \eta ^{0}\eta ^{1}$. The phase factor $\exp
\left( i\delta _{C}\right) $ is equal to $\pm 1$, thus $E\rightarrow -E$.
Note also that $J^{\mu }\rightarrow -J^{\mu }$, as should be expected for a
charge current. Meanwhile $C$ anticommutes with $[P,\beta ^{\mu }]$ and the
charge-conjugation operation entails no change on $A_{\mu }^{\left( 2\right)
}$. By the same token it can be shown that $A_{\mu }^{\left( 1\right) }$ and
$A_{\mu }^{\left( 2\right) }$ have opposite behavior under the time-reversal
transformation in such a way that both sorts of vector potentials change
sign under $\mathcal{PCT}$. The invariance of the nonminimal vector
potential under charge conjugation means that it does not couple to the
charge of the boson. In other words, $A_{\mu }^{\left( 2\right) }$ does not
distinguish particles from antiparticles. Hence, whether one considers
spin-0 or spin-1 bosons, this sort of interaction can not exhibit Klein's
paradox.

\subsection{Scalar sector}

For the case of spin 0, we use the representation for the $\beta ^{\mu }$\
matrices given by \cite{ned1}%
\begin{equation}
\beta ^{0}=%
\begin{pmatrix}
\theta & \overline{0} \\
\overline{0}^{T} & \mathbf{0}%
\end{pmatrix}%
,\quad \quad \beta ^{i}=%
\begin{pmatrix}
\widetilde{0} & \rho _{i} \\
-\rho _{i}^{T} & \mathbf{0}%
\end{pmatrix}%
,\quad i=1,2,3  \label{rep}
\end{equation}%
\noindent where%
\begin{eqnarray}
\ \theta &=&%
\begin{pmatrix}
0 & 1 \\
1 & 0%
\end{pmatrix}%
,\quad \quad \rho _{1}=%
\begin{pmatrix}
-1 & 0 & 0 \\
0 & 0 & 0%
\end{pmatrix}
\notag \\
&&  \label{rep2} \\
\rho _{2} &=&%
\begin{pmatrix}
0 & -1 & 0 \\
0 & 0 & 0%
\end{pmatrix}%
,\quad \quad \rho _{3}=%
\begin{pmatrix}
0 & 0 & -1 \\
0 & 0 & 0%
\end{pmatrix}
\notag
\end{eqnarray}%
\noindent $\overline{0}$, $\widetilde{0}$ and $\mathbf{0}$ are 2$\times $3, 2%
$\times $2 \ and 3$\times $3 zero matrices, respectively, while the
superscript T designates matrix transposition. Here the projection operator
can be written as \cite{gue}
\begin{equation}
P=\,\frac{1}{3}\left( \beta ^{\mu }\beta _{\mu }-1\right) =\text{diag}%
\,(1,0,0,0,0)  \label{proj}
\end{equation}%
In this case $P$ picks out the first component of the DKP spinor. The
five-component spinor can be written as $\psi ^{T}=\left( \psi _{1},...,\psi
_{5}\right) $ in such a way that the DKP equation for a boson constrained to
move along the $X$-axis decomposes into
\begin{equation*}
\left( D_{0}^{\left( -\right) }D_{0}^{\left( +\right) }-D_{1}^{\left(
-\right) }D_{1}^{\left( +\right) }+m^{2}\right) \psi _{1}=0
\end{equation*}%
\begin{equation}
D_{0}^{\left( +\right) }\psi _{1}=-im\psi _{2},\quad D_{1}^{\left( +\right)
}\psi _{1}=-im\psi _{3}  \label{DKP3}
\end{equation}%
\begin{equation*}
\psi _{4}=\psi _{5}=0
\end{equation*}%
where%
\begin{equation}
D_{\mu }^{\left( \pm \right) }=\partial _{\mu }+iA_{\mu }^{\left( 1\right)
}\pm A_{\mu }^{\left( 2\right) }  \label{dzao}
\end{equation}%
Furthermore,%
\begin{eqnarray}
J^{0} &=&\text{Re}\left( \psi _{2}^{\ast }\psi _{1}\right) =-\frac{1}{m}%
\text{Im}\left( \psi _{1}^{\ast }D_{0}^{\left( +\right) }\psi _{1}\right)
\notag \\
&&  \notag \\
J^{1} &=&-\,\text{Re}\left( \psi _{3}^{\ast }\psi _{1}\right) =\frac{1}{m}%
\text{Im}\left( \psi _{1}^{\ast }D_{1}^{\left( +\right) }\psi _{1}\right)
\label{corrente3} \\
&&  \notag \\
J^{2} &=&J^{3}=0  \notag
\end{eqnarray}%
Note that, in the absence of the nonminimal potential, the first line of (%
\ref{DKP3}) reduces to the Klein-Gordon equation, and that $\psi _{3}$, $%
\psi _{4}$ and $\psi _{5}$ are the superfluous components of the DKP spinor
(the reason that $\psi _{4}=\psi _{5}=0$ is because of the one-dimensional
movement).

In the time-independent case one has%
\begin{equation*}
\left( \frac{d^{2}}{dx^{2}}+2iA_{1}^{\left( 1\right) }\,\frac{d}{dx}%
+k^{2}\right) \phi _{1}=0
\end{equation*}%
\begin{equation}
\phi _{2}=\frac{1}{m}\left( E-A_{0}^{\left( 1\right) }+iA_{0}^{\left(
2\right) }\right) \,\phi _{1}  \label{dkp4}
\end{equation}%
\begin{equation*}
\phi _{3}=\frac{i}{m}\left( \frac{d}{dx}+iA_{1}^{\left( 1\right)
}+A_{1}^{\left( 2\right) }\right) \phi _{1}
\end{equation*}%
where%
\begin{equation*}
k^{2}=\left( E-A_{0}^{\left( 1\right) }\right) ^{2}-m^{2}-\left(
A_{1}^{\left( 1\right) }\right) ^{2}+i\frac{dA_{1}^{\left( 1\right) }}{dx}
\end{equation*}%
\begin{equation}
+\left( A_{0}^{\left( 2\right) }\right) ^{2}-\left( A_{1}^{\left( 2\right)
}\right) ^{2}+\frac{dA_{1}^{\left( 2\right) }}{dx}
\end{equation}%
Meanwhile,
\begin{equation}
J^{0}=\frac{E-A_{0}^{\left( 1\right) }}{m}\,|\phi _{1}|^{2},\quad J^{1}=%
\frac{1}{m}\left[ A_{1}^{\left( 1\right) }|\phi _{1}|^{2}+\text{Im}\left(
\phi _{1}^{\ast }\,\frac{d\phi _{1}}{dx}\right) \right]  \label{corrente4}
\end{equation}%
It is worthwhile to note that $J^{0}$ becomes negative in regions of space
where $E<A_{0}^{\left( 1\right) }$ (a circumstance associated to Klein's
paradox) and that $A_{\mu }^{\left( 2\right) }$ does not intervene
explicitly in the current. The orthonormalization formula (\ref{orto8})
becomes%
\begin{equation}
\int\limits_{-\infty }^{+\infty }dx\,\,\frac{\frac{E_{k}+E_{k^{\prime }}}{2}%
-A_{0}^{\left( 1\right) }}{m}\,\phi _{1k}^{\ast }\phi _{1k^{\prime }}=\pm
\delta _{E_{k}E_{k^{\prime }}}  \label{ORTO1}
\end{equation}%
regardless $A_{1}^{\left( 1\right) }$ and $A_{\mu }^{\left( 2\right) }$. Eq.
(\ref{ORTO1}) is in agreement with the orthonormalization formula for the
Klein-Gordon theory in the presence of a minimally coupled potential \cite%
{puk}. This is not surprising, because, after all, both DKP equation and
Klein-Gordon equation are equivalent under minimal coupling.

The form $\partial _{1}+iA_{1}^{\left( 1\right) }$ in Eq. (\ref{DKP3})
suggests that the space component of the minimal vector potential can be
gauged away by defining a new spinor
\begin{equation}
\tilde{\psi}=\exp \left( i\Lambda \right) \psi ,\quad A_{1}^{\left( 1\right)
}=\partial _{1}\Lambda
\end{equation}%
even if $A_{1}^{\left( 1\right) }$ is time dependent. Without any question%
\begin{equation}
\left( \partial _{1}+iA_{1}^{\left( 1\right) }\right) \psi =\exp \left(
-i\Lambda \right) \,\partial _{1}\tilde{\psi}  \label{300}
\end{equation}%
in such a way that $\tilde{\psi}$ satisfies the DKP equation without $%
A_{1}^{\left( 1\right) }$. In Refs. \cite{cla1} and \cite{kal} the term
involving $A_{1}^{\left( 2\right) }$ was explicitly absorbed into the wave
function. Nevertheless, it seems that there is no chance to get rid from
this term. As a matter of fact, we will show that the space component of the
nonminimal vector potential plays a peremptory role for confining bosons.
The possibility for ruling out \ $A_{1}^{\left( 1\right) }$ but not $%
A_{1}^{\left( 2\right) }$ is reinforced by the observation that the first
derivative of a second-order differential equation, such as the term
containing $A_{1}^{\left( 1\right) }$ in the first line of Eq. (\ref{dkp4}),
is a well-known trick in mathematics.

It is noticeable that if $|A_{\mu }^{\left( 2\right) }|\rightarrow \infty $
as $|x|\rightarrow \pm \infty $, confining solutions for a pure nonminimal
vector potential will be possible on the condition that the space component
of $A_{\mu }^{\left( 2\right) }$ is stronger, or has a dominant asymptotic
behavior, than its time component. Otherwise, nothing but continuum states
will be possible. In this last circumstance, a boson can tunnel into the
classically forbidden region, an unexpected result in nonrelativistic
mechanics and by no means related to Klein's paradox. On the other hand, for
a pure nonminimal vector potential going to zero at infinity, a necessary
condition for the existence of bound-state solutions (with $|E|<m$) is that
\begin{equation}
\left( A_{0}^{\left( 2\right) }\right) ^{2}-\left( A_{1}^{\left( 2\right)
}\right) ^{2}+\frac{dA_{1}^{\left( 2\right) }}{dx}>0
\end{equation}%
at any arbitrary point on the $X$-axis. In this case, it is the time
component of the nonminimal vector potential that plays a leading role in
establishing bound states.

\subsection{ Vector sector}

For the case of spin 1, the $\beta ^{\mu }$\ matrices are \cite{ned2}%
\begin{equation}
\beta ^{0}=%
\begin{pmatrix}
0 & \overline{0} & \overline{0} & \overline{0} \\
\overline{0}^{T} & \mathbf{0} & \mathbf{I} & \mathbf{0} \\
\overline{0}^{T} & \mathbf{I} & \mathbf{0} & \mathbf{0} \\
\overline{0}^{T} & \mathbf{0} & \mathbf{0} & \mathbf{0}%
\end{pmatrix}%
,\quad \quad \beta ^{i}=%
\begin{pmatrix}
0 & \overline{0} & e_{i} & \overline{0} \\
\overline{0}^{T} & \mathbf{0} & \mathbf{0} & -is_{i} \\
-e_{i}^{T} & \mathbf{0} & \mathbf{0} & \mathbf{0} \\
\overline{0}^{T} & -is_{i} & \mathbf{0} & \mathbf{0}%
\end{pmatrix}
\label{betaspin1}
\end{equation}%
\noindent where $s_{i}$ are the 3$\times $3 spin-1 matrices $\left(
s_{i}\right) _{jk}=-i\varepsilon _{ijk}$, $e_{i}$ are the 1$\times $3
matrices $\left( e_{i}\right) _{1j}=\delta _{ij}$ and $\overline{0}=%
\begin{pmatrix}
0 & 0 & 0%
\end{pmatrix}%
$, while\textbf{\ }$\mathbf{I}$ and $\mathbf{0}$\textbf{\ }designate the 3$%
\times $3 unit and zero matrices, respectively. In this representation
\begin{equation}
P=\,\beta ^{\mu }\beta _{\mu }-2=\text{diag}\,(1,1,1,1,0,0,0,0,0,0)
\label{p1}
\end{equation}%
\textit{i.e.}, $P$ projects out the four upper components of the DKP spinor.
\noindent With the spinor written as $\psi ^{T}=\left( \psi _{1},...,\psi
_{10}\right) $, and partitioned as%
\begin{equation*}
\psi _{I}^{\left( +\right) }=\left(
\begin{array}{c}
\psi _{3} \\
\psi _{4}%
\end{array}%
\right) ,\quad \psi _{I}^{\left( -\right) }=\psi _{5}
\end{equation*}%
\begin{equation}
\psi _{II}^{\left( +\right) }=\left(
\begin{array}{c}
\psi _{6} \\
\psi _{7}%
\end{array}%
\right) ,\quad \psi _{II}^{\left( -\right) }=\psi _{2}  \label{part}
\end{equation}%
\begin{equation*}
\psi _{III}^{\left( +\right) }=\left(
\begin{array}{c}
\psi _{10} \\
-\psi _{9}%
\end{array}%
\right) ,\quad \psi _{III}^{\left( -\right) }=\psi _{1}
\end{equation*}%
the one-dimensional DKP equation can be expressed in the compact form
\begin{equation*}
\left( D_{0}^{\left( \mp \right) }D_{0}^{\left( \pm \right) }-D_{1}^{\left(
\mp \right) }D_{1}^{\left( \pm \right) }+m^{2}\right) \psi _{I}^{\left( \pm
\right) }=0
\end{equation*}%
\begin{equation}
D_{0}^{\left( \pm \right) }\psi _{I}^{\left( \pm \right) }=-im\psi
_{II}^{\left( \pm \right) },\quad D_{1}^{\left( \pm \right) }\psi
_{I}^{\left( \pm \right) }=-im\psi _{III}^{\left( \pm \right) }  \label{DKp3}
\end{equation}%
\begin{equation*}
\psi _{8}=0
\end{equation*}%
where $D_{\mu }^{\left( \pm \right) }$ is again given by (\ref{dzao}). In
addition, expressed in terms of (\ref{part}) the current can be written as
\begin{equation*}
J^{0}=\text{Re}\left( \psi _{II}^{\left( +\right) \dagger }\psi _{I}^{\left(
+\right) }+\psi _{II}^{\left( -\right) \dagger }\psi _{I}^{\left( -\right)
}\right) =-\frac{1}{m}\,\text{Im}\left( \psi _{I}^{\left( +\right) \dagger
}D_{0}^{\left( +\right) }\psi _{I}^{\left( +\right) }+\psi _{I}^{\left(
-\right) \dagger }D_{0}^{\left( -\right) }\psi _{I}^{\left( -\right) }\right)
\end{equation*}%
\begin{equation}
J^{1}=-\text{Re}\left( \psi _{III}^{\left( +\right) \dagger }\psi
_{I}^{\left( +\right) }+\psi _{III}^{\left( -\right) \dagger }\psi
_{I}^{\left( -\right) }\right) =\frac{1}{m}\,\text{Im}\left( \psi
_{I}^{\left( +\right) \dagger }D_{1}^{\left( +\right) }\psi _{I}^{\left(
+\right) }+\psi _{I}^{\left( -\right) \dagger }D_{1}^{\left( -\right) }\psi
_{I}^{\left( -\right) }\right)  \label{CUR}
\end{equation}%
\begin{equation*}
J^{2}=J^{3}=0
\end{equation*}%
Note that the third line plus the second equation in the middle line of (\ref%
{DKp3}) are the constraint equations which allow one to eliminate the
superfluous components ($\psi _{1}$, $\psi _{8}$, $\psi _{9}$ and $\psi
_{10} $) of the DKP spinor. The component $\psi _{8}=0$ because the movement
is restrict to the $X$-axis.

Meanwhile the time-independent DKP equation decomposes into%
\begin{equation*}
\left( \frac{d^{2}}{dx^{2}}+2iA_{1}^{\left( 1\right) }\,\frac{d}{dx}+k_{\pm
}^{2}\right) \phi _{I}^{\left( \pm \right) }=0
\end{equation*}%
\begin{equation}
\phi _{II}^{\left( \pm \right) }=\frac{1}{m}\left( E-A_{0}^{\left( 1\right)
}\pm iA_{0}^{\left( 2\right) }\right) \,\phi _{I}^{\left( \pm \right) }
\label{spin1-ti}
\end{equation}%
\begin{equation*}
\phi _{III}^{\left( \pm \right) }=\frac{i}{m}\left( \frac{d}{dx}%
+iA_{1}^{\left( 1\right) }\pm A_{1}^{\left( 2\right) }\right) \phi
_{I}^{\left( \pm \right) }
\end{equation*}%
where%
\begin{equation*}
k_{\pm }^{2}=\left( E-A_{0}^{\left( 1\right) }\right) ^{2}-m^{2}-\left(
A_{1}^{\left( 1\right) }\right) ^{2}+i\frac{dA_{1}^{\left( 1\right) }}{dx}
\end{equation*}%
\begin{equation}
+\left( A_{0}^{\left( 2\right) }\right) ^{2}-\left( A_{1}^{\left( 2\right)
}\right) ^{2}\pm \frac{dA_{1}^{\left( 2\right) }}{dx}  \label{K}
\end{equation}%
Now the components of the four-current are
\begin{equation*}
J^{0}=\frac{E-A_{0}^{\left( 1\right) }}{m}\left( |\phi _{I}^{\left( +\right)
}|^{2}+|\phi _{I}^{\left( -\right) }|^{2}\right)
\end{equation*}%
\begin{equation}
J^{1}=\frac{1}{m}\left[ A_{1}^{\left( 1\right) }\left( |\phi _{I}^{\left(
+\right) }|^{2}+|\phi _{I}^{\left( -\right) }|^{2}\right) +\text{Im}\left(
\phi _{I}^{\left( +\right) \dagger }\,\frac{d\phi _{I}^{\left( +\right) }}{dx%
}+\phi _{I}^{\left( -\right) \dagger }\,\frac{d\phi _{I}^{\left( -\right) }}{%
dx}\right) \right]  \label{CUR2}
\end{equation}%
and the orthonormalization expression (\ref{orto8}) takes the form
\begin{equation}
\int\limits_{-\infty }^{+\infty }dx\,\,\frac{\frac{E_{k}+E_{k^{\prime }}}{2}%
-A_{0}^{\left( 1\right) }}{m}\,\left( \phi _{Ik}^{\left( +\right) \dagger
}\phi _{Ik^{\prime }}^{\left( +\right) }+\phi _{Ik}^{\left( -\right) \dagger
}\phi _{Ik^{\prime }}^{\left( -\right) }\right) =\pm \delta
_{E_{k}E_{k^{\prime }}}
\end{equation}%
Just as for scalar bosons, $J^{0}<0$ for $E<A_{0}^{\left( 1\right) }$ and $%
A_{\mu }^{\left( 2\right) }$ does not appear in the current. Similarly, $%
A_{1}^{\left( 1\right) }$ and $A_{\mu }^{\left( 2\right) }$ do not manifest
explicitly in the orthonormalization formula.

From (\ref{spin1-ti})-(\ref{K}) one sees that the solution for the spin-1
sector consists in searching solutions for two Klein-Gordon-like equations,
owing to the term $dA_{1}^{\left( 2\right) }/dx$ in (\ref{K}). It should not
be forgotten, though, that the equations for $\phi _{I}^{\left( +\right) }$
and $\phi _{I}^{\left( -\right) }$ are not indeed independent because $E$
appears in both equations. Evidently, matching a common value for the energy
might compromise the existence of solutions for spin-1 bosons when compared
to the solutions for spin-0 bosons with the very same potentials. This
amounts to say that the solutions for the spin-1 sector of the DKP theory,
if they really exist, can be obtained from a restrict class of solutions of
the spin-0 sector. This limitation on the possible solutions for spin-1
bosons as compared for spin-0 bosons should not be a surprise if one
remembers that, in the absence of any interaction, all the components of the
free Proca equation obey a free Klein-Gordon equation but with an additional
constraint on the components of the Proca field.

\section{The nonminimal vector linear potential}

Having set up the spin-0 and spin-1 equations for vector interactions, we
are now in a position to use the machinery developed above in order to solve
the DKP equation with specific forms for nonminimal interactions. Let us
consider pure nonminimal vector linear potentials in the form%
\begin{equation}
A_{0}^{\left( 2\right) }=m^{2}\omega _{0}|x|,\quad A_{1}^{\left( 2\right)
}=m^{2}\omega _{1}x,  \label{pot2}
\end{equation}%
where $\omega _{0}$ and $\omega _{1}$ are dimensionless quantities. Our
problem is to solve (\ref{dkp4}) and (\ref{spin1-ti}) for $\phi $ and to
determine the allowed energies. Although the absolute value of $x$ in $%
A_{0}^{\left( 2\right) }$ is irrelevant in the effective equations for $\phi
_{1}$ (in the scalar sector) and $\phi _{I}^{\left( \pm \right) }$ (in the
vector sector), it is there for ensuring the covariance of the DKP theory
under parity. It follows that the DKP spinor will have a definite parity and
$A^{\mu }$ and $J^{\mu }$ will be genuine four-vectors.

\subsection{Scalar sector}

\begin{figure}[th]
\begin{center}
\includegraphics[width=9cm, angle=0]{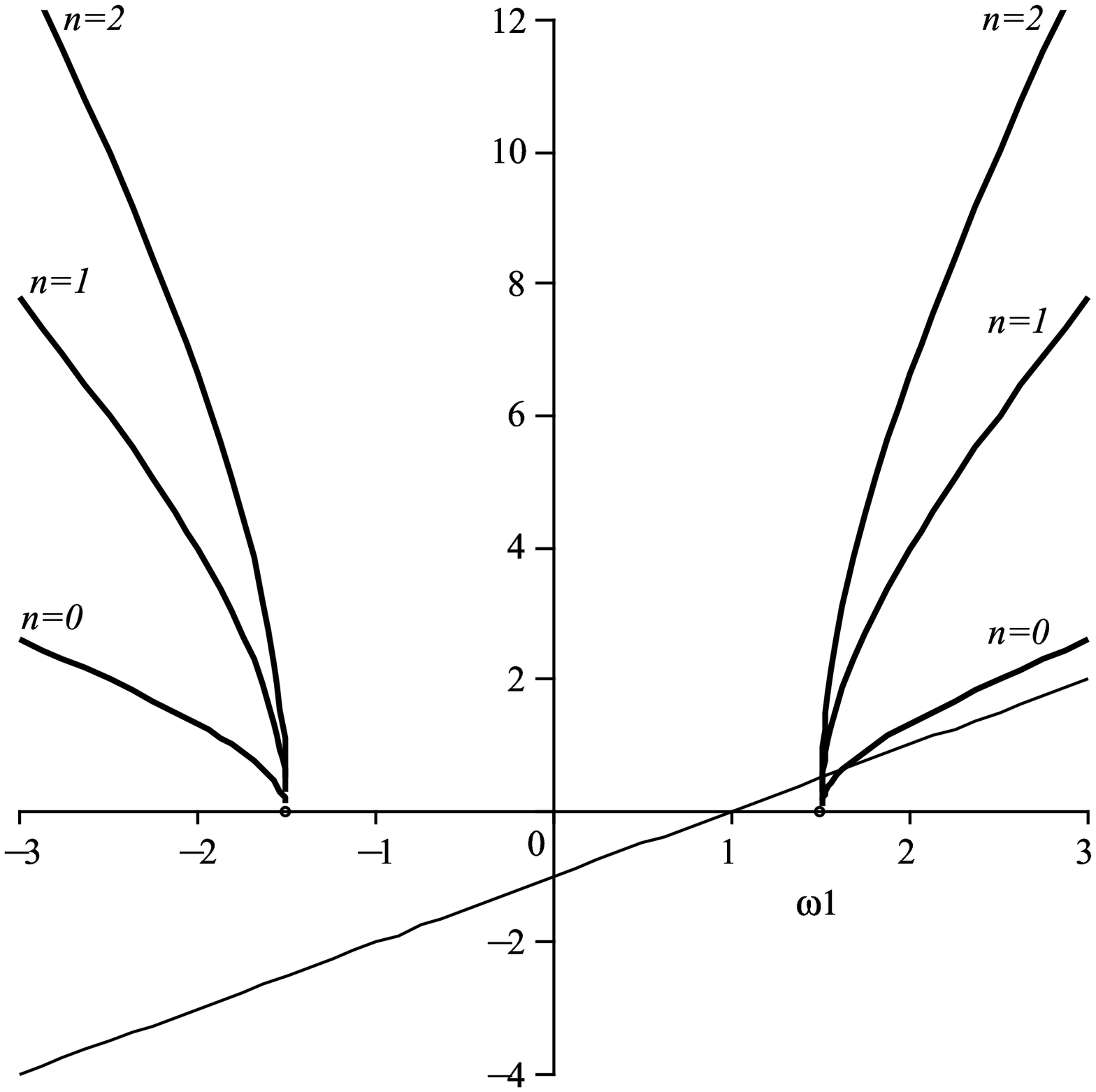}
\end{center}
\par
\vspace*{-0.1cm}
\caption{Graphical solution of (\protect\ref{const}) for $|\protect\omega %
_{0}|=1.5$ for the three lowest quantum numbers. Heavy lines for $f_{H}(%
\protect\omega _{1})$ and light line for $f_{S}(\protect\omega _{1})$.}
\label{Fig1}
\end{figure}

For the spin-0 sector of the DKP theory one finds that $\phi _{1}$ obeys the
second-order differential equation
\begin{equation}
\frac{d^{2}\phi _{1}}{dx^{2}}+\left( \varepsilon ^{2}-m^{4}\Omega
^{2}x^{2}\right) \phi _{1}=0  \label{ohs1}
\end{equation}%
where%
\begin{equation}
\varepsilon ^{2}=E^{2}-m^{2}+m^{2}\omega _{1},\quad \Omega ^{2}=\omega
_{1}^{2}-\omega _{0}^{2}  \label{ohs2}
\end{equation}%
The solution for (\ref{ohs1}), with $\varepsilon ^{2}>0$ and $\Omega ^{2}>0$%
, is precisely the well-known solution of the Schr\"{o}dinger equation for
the nonrelativistic harmonic oscillator (see, \textit{e.g.}, \cite{sch}):%
\begin{equation}
\varepsilon _{n}^{2}=\left( 2n+1\right) m^{2}|\Omega |  \label{eigen1}
\end{equation}%
\begin{equation}
\left( \phi _{1}\right) _{n}=N_{n}H_{n}\left( m\sqrt{|\Omega |}x\right) \exp
\left( -\frac{m^{2}|\Omega |}{2}x^{2}\right)  \label{wf}
\end{equation}%
where $n=0,1,2,\ldots $, $N_{n}$ is a normalization constant, and $%
H_{n}\left( \zeta \right) $ is a $n$-th degree Hermite polynomial in $\zeta $%
. Notice that the condition $\Omega ^{2}>0$ requires that $|\omega
_{1}|>|\omega _{0}|$, meaning that the space component of the potential must
be stronger than its time component in order to the effective potential be a
true confining potential. Nevertheless, there is no requirement on the signs
of $\omega _{1}$ and $\omega _{0}$. From (\ref{eigen1}) one obtains the
discrete set of DKP energies (symmetrical about $E=0$ as it should be since $%
A_{\mu }^{\left( 2\right) }$ does not distinguish particles from
antiparticles) $E_{n}=\pm |E_{n}|$ where
\begin{equation}
|E_{n}|=m\sqrt{1-\omega _{1}+\left( 2n+1\right) |\Omega |}  \label{eigen2}
\end{equation}%
irrespective to the sign of $\omega _{0}$. In general, $|E_{n}|$ is higher
for $\omega _{1}<0$ than for $\omega _{1}>0$. It increases with the quantum
number and it is a monotonically decreasing function of $\omega _{0}$. In
order to insure the reality of the spectrum, the coupling constants $\omega
_{0}$ and $\omega _{1}$ satisfy the additional constraint%
\begin{equation}
\left( 2n+1\right) \sqrt{\omega _{1}^{2}-\omega _{0}^{2}}\geqslant \omega
_{1}-1  \label{const}
\end{equation}%
If one squares (\ref{const}) the resulting inequality is in general a
quadratic algebraic inequality in $\omega _{1}$ (or $\omega _{0}$), which
can be solved analytically. The price paid is that some spurious solutions
can appear in this process, although, of course, these can be eliminated by
checking whether they satisfy the original inequality. A more instructive
procedure is to follow a graphical method, by which one seeks the regions of
the functions of $\omega _{1}$ in (\ref{const}): a hyperbole on the
left-hand side,%
\begin{equation}
f_{H}\left( \omega _{1}\right) =\left( 2n+1\right) \sqrt{\omega
_{1}^{2}-\omega _{0}^{2}}  \label{fh}
\end{equation}%
and a straight line on the right-hand side,%
\begin{equation}
f_{S}\left( \omega _{1}\right) =\omega _{1}-1  \label{fs}
\end{equation}%
$f_{H}\left( \omega _{1}\right) $ is a nonnegative function having two
symmetric branches, and for $|\omega _{1}|\gg |\omega _{0}|$ it approximates
the function $\left( 2n+1\right) |\omega _{1}|$. Figure \ref{Fig1} present
results for the three first quantum numbers with $|\omega _{0}|>1$. For $%
\omega _{1}>|\omega _{0}|$, this figure shows clearly that $f_{H}\geqslant
f_{S}$ only for some $\omega _{1}\geqslant \left( \tilde{\omega}_{1}\right)
_{n}>|\omega _{0}|$, although $f_{H}>f_{S}$ for all $\omega _{1}<-|\omega
_{0}|$. The intersection points of $f_{H}$ and $f_{S}$, for $|\omega _{0}|>1$%
, correspond to $|E_{n}|=0.$ Figure \ref{Fig1} also allows one to conclude
that $|E_{n}|>0$ for $|\omega _{0}|<1$. Notice that there is a high density
of states (number of states in a fixed range of energy) corresponding to an
infinite set of quasi-degenerate solutions in the neighborhood of $|\omega
_{1}|=|\omega _{0}|$.

In the weak-coupling limit, $\omega _{1}\ll 1$ and $|\Omega |\ll 1$, $%
|E_{n}|\simeq m$ for small quantum numbers, and (\ref{eigen2}) becomes%
\begin{equation}
|E_{n}|\simeq m\left[ 1-\frac{\omega _{1}}{2}+\left( n+\frac{1}{2}\right)
|\Omega |\right]  \label{nr}
\end{equation}%
This equally spaced energy spectrum is a sort of nonrelativistic limit.
Therefore, it can be said that the linear potentials given by (\ref{pot2})
describe a genuine nonminimal vector DKP oscillator. Nevertheless, the
Lorentz structure of the potentials plays no role in a nonrelativistic
scheme, because one has to use the Schr\"{o}dinger equation with the
potential $A_{0}^{\left( 2\right) }+A_{1}^{\left( 2\right) }$. Despite the
effective harmonic oscillator potential appearing in (\ref{ohs1}), the
linear potentials \ given by (\ref{pot2}) do not furnish bound-state
solutions in the Schr\"{o}dinger equation because the sum $A_{0}^{\left(
2\right) }+A_{1}^{\left( 2\right) }$ with $A_{1}^{\left( 2\right) }\neq 0$
is unbounded from below.

On the other hand, for $|\omega _{1}|\gg |\omega _{0}|$ one has that%
\begin{equation}
|E_{n}|\simeq m\sqrt{1-\omega _{1}+\left( 2n+1\right) |\omega _{1}|}
\label{lim2}
\end{equation}%
so that $|E_{n}|>m$ for $\omega _{1}<0$. Concerning $\omega _{1}>0$, as far
as $\omega _{1}$ increases, the spectrum moves towards $E=0$, except for $%
\omega _{0}=0$ which maintains $|E_{n}|\geq m$ (the spectrum acquiesces $%
|E_{0}|=m$ in this limit case).

\begin{figure}[th]
\begin{center}
\includegraphics[width=9cm, angle=0]{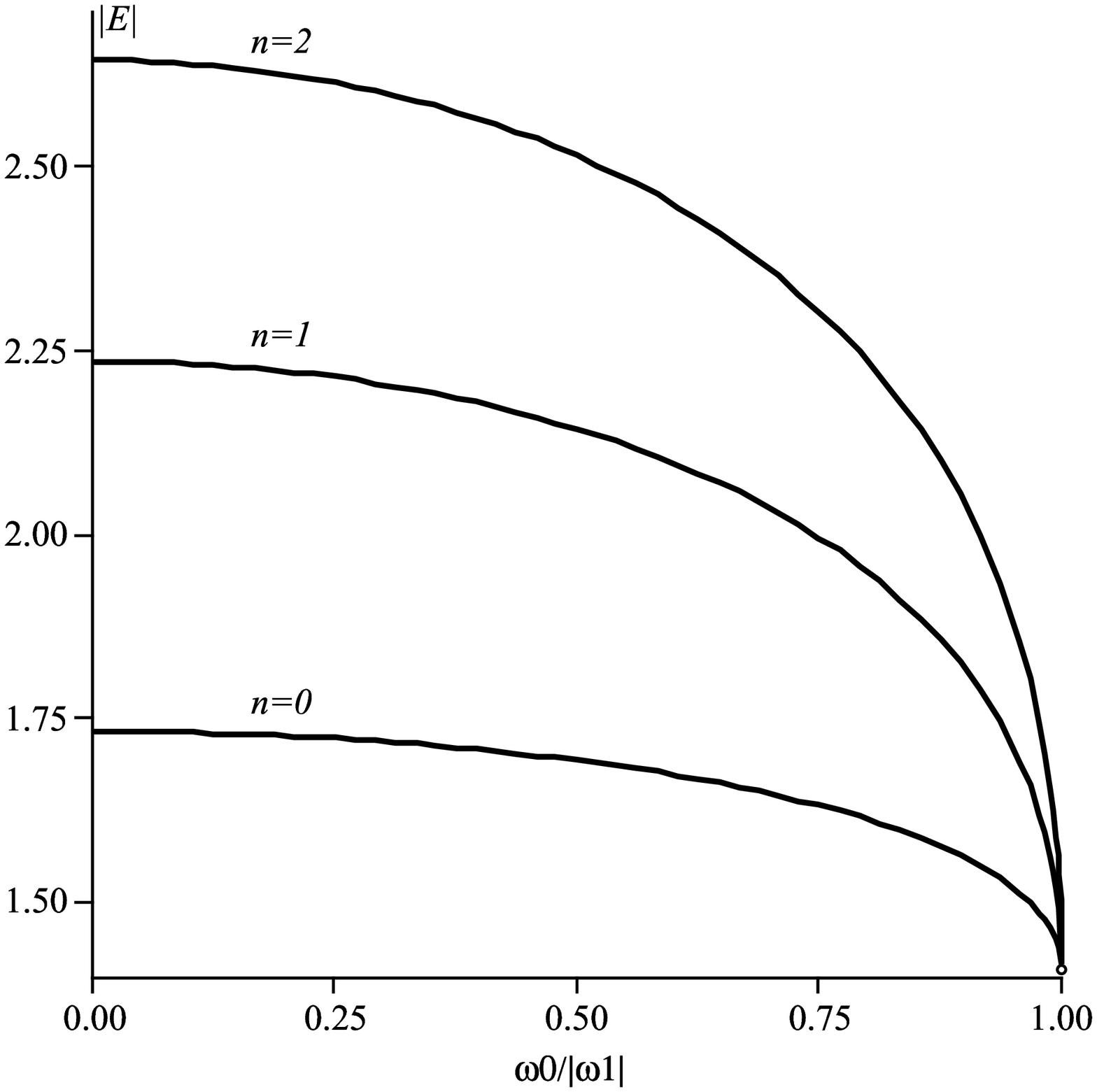}
\end{center}
\par
\vspace*{-0.1cm}
\caption{Positive spectrum of spin-0 bosons for the three lowest quantum
numbers as a function of $\protect\omega _{0}/|\protect\omega _{1}|$, for $%
\protect\omega _{1}=-1$ ($m=1$).}
\label{Fig2}
\end{figure}

\begin{figure}[th]
\begin{center}
\includegraphics[width=9cm, angle=0]{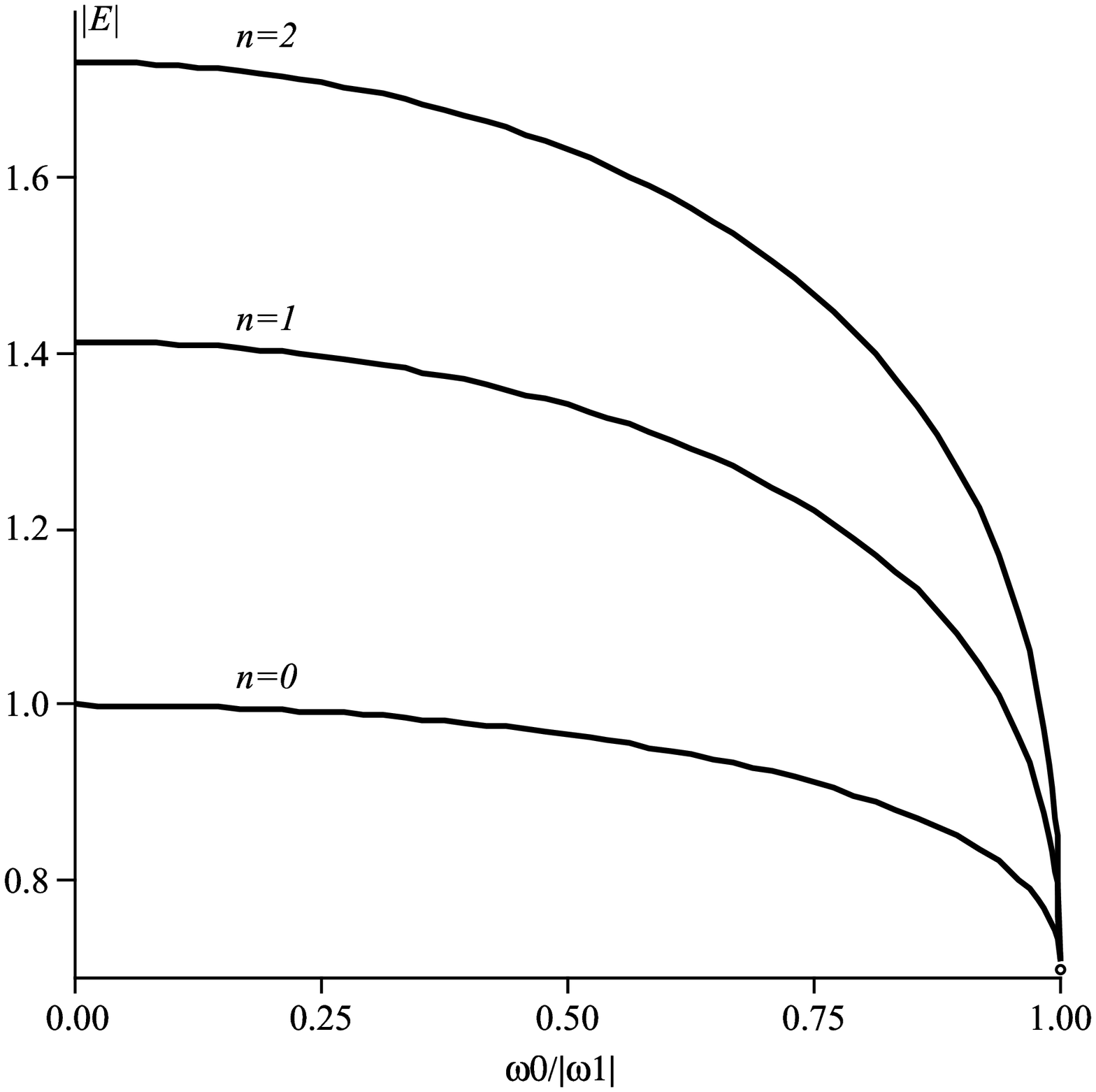}
\end{center}
\par
\vspace*{-0.1cm}
\caption{The same as Figure 2, for $\protect\omega _{1}=0.5$.}
\label{Fig3}
\end{figure}

\begin{figure}[th]
\begin{center}
\includegraphics[width=9cm, angle=0]{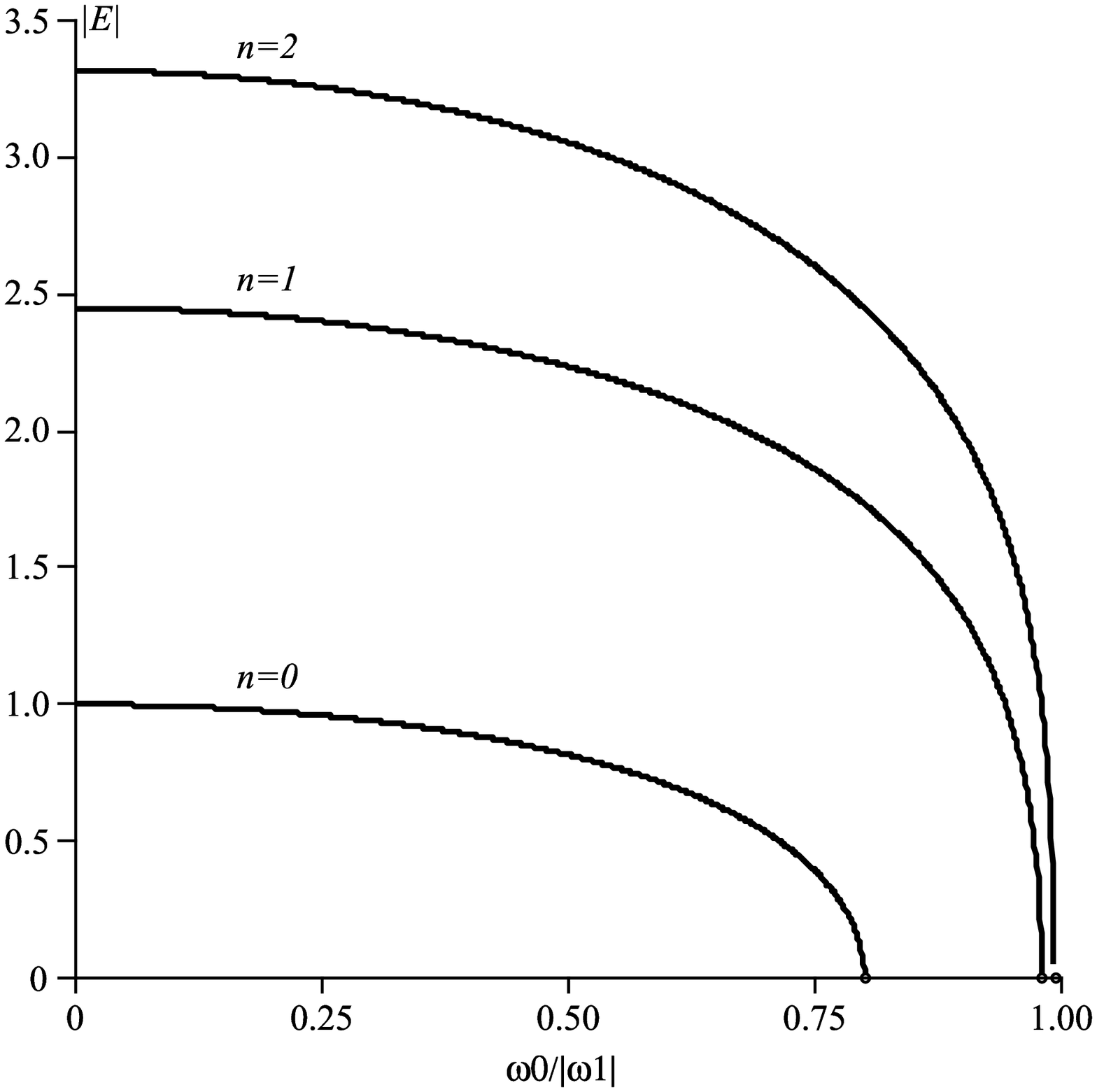}
\end{center}
\par
\vspace*{-0.1cm}
\caption{The same as Figure 2, for $\protect\omega _{1}=2.5$.}
\label{Fig4}
\end{figure}

\begin{figure}[th]
\begin{center}
\includegraphics[width=9cm, angle=0]{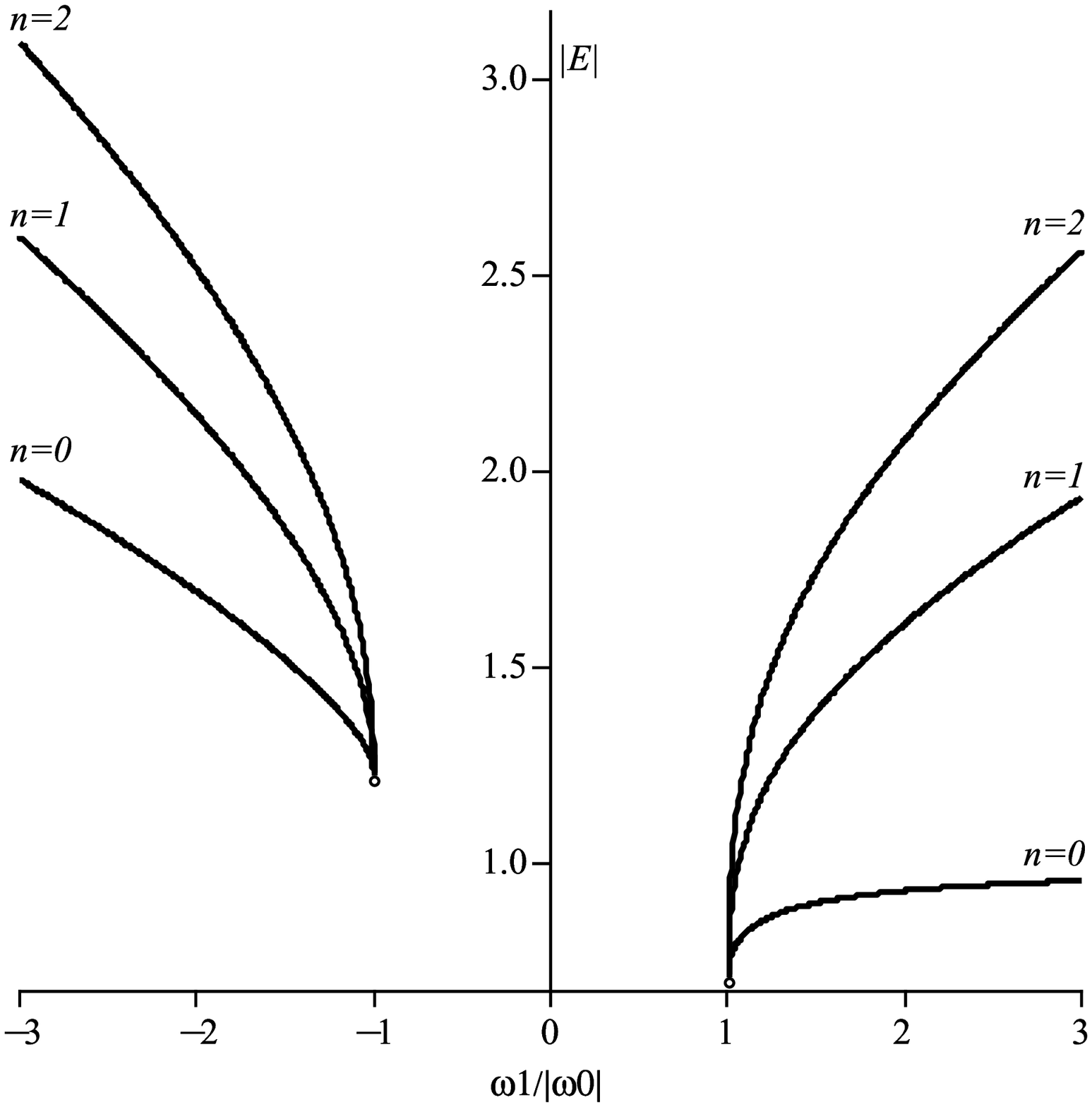}
\end{center}
\par
\vspace*{-0.1cm}
\caption{Positive spectrum of spin-0 bosons for the three lowest quantum
numbers as a function $\protect\omega _{1}/\protect\omega _{0}$, for $%
\protect\omega _{0}=0.5$ ($m=1 $).}
\label{Fig5}
\end{figure}

\begin{figure}[th]
\begin{center}
\includegraphics[width=9cm, angle=0]{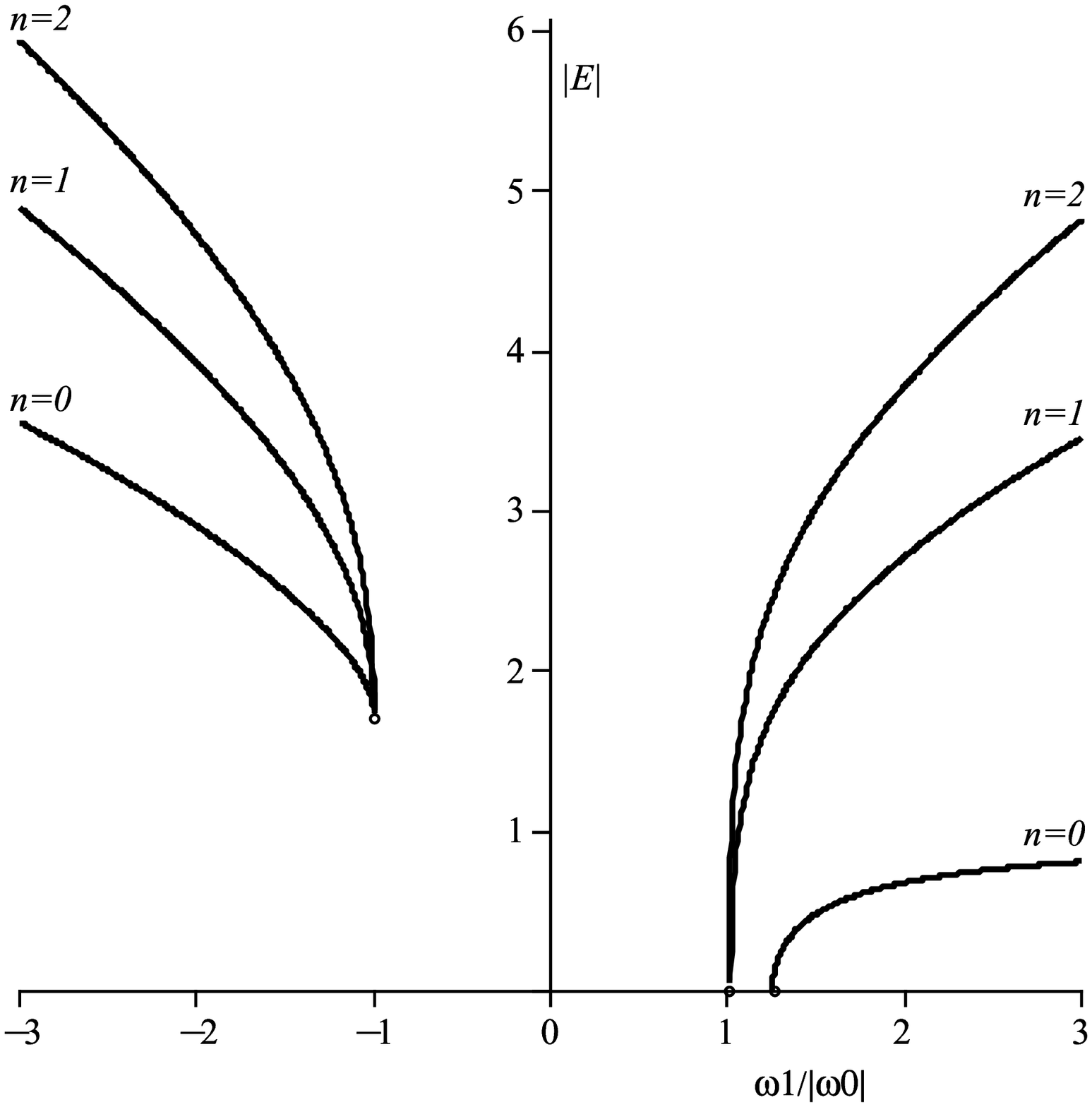}
\end{center}
\par
\vspace*{-0.1cm}
\caption{The same as Figure 5, for $\protect\omega _{0}=2$.}
\label{Fig6}
\end{figure}

Figures \ref{Fig2}, \ref{Fig3} and \ref{Fig4} illustrate the spectrum in
terms of $\omega _{0}/|\omega _{1}|$ for three different values of $\omega
_{1}$. For $\omega _{1}<1$ there is a spectral gap given by
\begin{equation}
2m\sqrt{1-\omega _{1}+\left( 2n+1\right) |\Omega |}  \label{gap}
\end{equation}%
and there are infinitely many energy levels above $m$ where, in the absence
of interaction, there was the continuum. As far as $\omega _{1}$ increases,
the spectrum moves towards $E=0$, except for $\omega _{0}=0$. The gap tends
to vanish as $\omega _{1}$ becomes close to $1$ (for $\omega _{0}\neq 0$),
and so the positive- and negative-energy levels tend to be very close to
each other. Figures \ref{Fig5} and \ref{Fig6} illustrate the spectrum in
terms of $\omega _{1}/\omega _{0}$ for two different values of $\omega _{0}$.

The charge density%
\begin{equation}
J^{0}=\frac{E}{m}\,|\phi _{1}|^{2}  \label{jta0}
\end{equation}%
dictates that $\phi _{1}$ must be normalized as%
\begin{equation}
\frac{|E|}{m}\int\limits_{-\infty }^{+\infty }dx\,|\phi _{1}|^{2}=1
\label{norma1}
\end{equation}%
Using the property \cite{sch}%
\begin{equation}
\int\limits_{-\infty }^{+\infty }d\zeta \,H_{n}^{2}\left( \zeta \right) \exp
\left( -\zeta ^{2}\right) =2^{n}n!\sqrt{\pi }  \label{herm}
\end{equation}%
one finds that the normalization constant can be chosen to be%
\begin{equation}
N_{n}=\left( \frac{m|\Omega |}{\pi }\right) ^{1/4}\sqrt{\frac{m}{%
2^{n}n!\,|E_{n}|}},\quad \mathrm{for}\quad E_{n}\neq 0  \label{norma2}
\end{equation}%
Thus, for $E_{n}\neq 0$ one has%
\begin{equation}
J_{n}^{0}(x)=\frac{\mathrm{sign}\left( E_{n}\right) }{2^{n}n!}\sqrt{\frac{%
m|\Omega |}{\pi }}H_{n}^{2}\left( m\sqrt{|\Omega |}x\right) \exp \left(
-m^{2}|\Omega |x^{2}\right)  \label{JOTA0}
\end{equation}%
Then, using (\ref{exp}) the quantity $\left( \Delta x_{n}\right)
^{2}=\left\langle x^{2}\right\rangle _{n}-\left\langle x\right\rangle
_{n}^{2}$ can be written as%
\begin{equation*}
\left( \Delta x_{n}\right) ^{2}=\int\limits_{-\infty }^{+\infty
}dx\,\left\vert J_{n}^{0}(x)\right\vert \,x^{2}-\left( \int\limits_{-\infty
}^{+\infty }dx\,\left\vert J_{n}^{0}(x)\right\vert \,x\right) ^{2}
\end{equation*}%
Now it is a simple matter to write down the uncertainty in the
position:\bigskip
\begin{equation}
\Delta x_{n}=\sqrt{\frac{n+1/2}{m^{2}|\Omega |}}  \label{dx}
\end{equation}%
If $\Delta x_{n}$ shrinks then $\Delta p_{n}$ (uncertainty in the momentum)
will must swell, in consonance with the Heisenberg uncertainty principle.
Nevertheless, the maximum uncertainty in the momentum is given by $m$
requiring that is impossible to localize a boson in a region of space less
than half of its Compton wavelength (see, for example, \cite{gre}-\cite{str}%
). Nevertheless, if one defines an effective mass as $m_{\mathtt{eff}}=m%
\sqrt{|\Omega |}$ and an effective Compton wavelength as $\lambda _{\mathtt{%
eff}}=1/m_{\mathtt{eff}}$ one will find that $\Delta x_{n}=\lambda _{\mathtt{%
eff}}\sqrt{n+1/2}$. It follows that the high localization of bosons, related
to high values of $|\Omega |$ ($|\omega _{1}|\gg |\omega _{0}|$) never
menaces the single-particle interpretation of the DKP theory. For $|\Omega
|\simeq 0$ ($|\omega _{1}|\simeq |\omega _{0}|$) one has that $m_{\mathtt{eff%
}}\simeq 0$ and the quasi-degenerate solutions mentioned above are related
to very delocalized states. As for the behavior in the neighborhood of $%
E_{n}=0$ one should note that, despite of $J_{n}^{0}$ and $\Delta x_{n}$ are
independent of $E_{n}$, the DKP spinor is not defined for $|E_{n}|=0$. Thus,
$E_{n}=0$ must be ruled out of the theory. Although positive- and
negative-energy levels do not touch, they can be very close to each other
for coupling constants moderately strong without any danger of reaching the
conditions for Klein's paradox.

\subsection{Vector sector}

As for the spin-1 sector, proceeding as before, one finds that $\phi
_{I}^{\left( \pm \right) }$ obeys the equation
\begin{equation}
\frac{d^{2}\phi _{I}^{\left( \pm \right) }}{dx^{2}}+\left( \varepsilon _{\pm
}^{2}-m^{4}\Omega ^{2}x^{2}\right) \phi _{I}^{\left( \pm \right) }=0
\end{equation}%
where $\Omega ^{2}$ is defined as in (\ref{ohs2}) and%
\begin{equation}
\varepsilon _{\pm }^{2}=E^{2}-m^{2}\pm m^{2}\omega _{1}  \label{spin1-eps}
\end{equation}%
For bound states, to which we shall devote our attention, we must require $%
\varepsilon _{\pm }^{2}>0$ and $\Omega ^{2}>0$, as before. Thus, the
solution is expressed as%
\begin{equation}
\varepsilon _{n_{\pm }}^{2}=\left( 2n_{\pm }+1\right) m^{2}|\Omega |
\label{spinEPS}
\end{equation}%
\begin{equation}
\left( \phi _{I}^{\left( \pm \right) }\right) _{n_{\pm }}=N_{n_{\pm
}}H_{n_{\pm }}\left( m\sqrt{|\Omega |}x\right) \exp \left( -\frac{%
m^{2}|\Omega |}{2}x^{2}\right)  \label{func}
\end{equation}%
where $n_{\pm }=0,1,2,\ldots $ , $N_{n_{-}}$ is a normalization constant,
and $N_{n_{+}}=\left( N_{3},N_{4}\right) ^{T}$ is a column matrix whose
elements are normalization constants related to the solutions for $\phi _{3}$
and $\phi _{4}$. Hence, the necessary conditions for binding spin-1 bosons
subject to linear potentials have been put forward. The formal analytical
solutions have been obtained and it has been revealed that the solutions
related to the spinor $\phi _{I}^{\left( +\right) }$ are formally the same
as those ones for spin-0 bosons. Now we move on to match a common energy to
the spin-1 boson problem. The matching condition requires that the quantum
numbers $n_{+}$ and $n_{-}$ must satisfy the relation%
\begin{equation}
n_{+}-n_{-}=\frac{1}{\sqrt{1-\left( \frac{\omega _{0}}{\omega _{1}}\right)
^{2}}}\frac{\omega _{1}}{|\omega _{1}|}  \label{vin}
\end{equation}%
This constraint on the nodal structure of $\phi _{I}^{\left( +\right) }$ and
$\phi _{I}^{\left( -\right) }$ dictates that acceptable solutions only occur
for a countable number of possibilities for $|\omega _{1}|/|\omega _{0}|$,
\textit{viz.}
\begin{equation}
\frac{|\omega _{0}|}{|\omega _{1}|}=\sqrt{1-\frac{1}{\left(
n_{+}-n_{-}\right) ^{2}}}
\end{equation}

\section{Conclusions}

We showed that minimal and nonminimal vector interactions behave differently
under charge-conjugation and time-reversal transformations. Although Klein's
paradox can not be treated as unworthy of regard in the DKP theory with
minimally coupled vector interactions, it never makes its appearance in the
case of nonminimal vector interactions because they do not couple to the
charge.

In the case of a pure nonminimal vector coupling, both particle and particle
energy levels are members of the spectrum, and the particle and antiparticle
spectra are symmetrical about $E=0$. If the interaction potential is
attractive (repulsive) for bosons it will also be attractive (repulsive) for
antibosons. However, there is no crossing of levels because possible states
in the strong field regime with $E=0$ are in fact unnormalizable. These
facts imply that there is no channel for spontaneous boson-antiboson
creation and for that reason the single-particle interpretation of the DKP
equation is ensured. The charge conjugation operation allows us to migrate
from the spectrum of particles to the spectrum of antiparticles and vice
versa just by changing the sign of $E$. This change induces no change in the
nodal structure of the components of the DKP spinor and so the nodal
structure of the four-current is preserved.

In view of recent developments on the construction of positive-definite
inner product for the Klein-Gordon theory \cite{mos}, we acknowledge that we
took a very conservative stance when considering a current that can not be
related to a probability current. The interesting possibility of a
probability current in the DKP theory, constructed from the energy-momentum
tensor, launched in \cite{gho1}-\cite{gho2} , though, received a severe
criticism in \cite{stru} and \cite{tati}. It will then be challenging to
construct a probability current in the DKP theory from a relativistically
invariant positive-definite inner product. Notwithstanding, the conserved
charge current plus the charge conjugation operation are enough to infer
about the absence of Klein's paradox under nonminimal vector interactions,
or its possible presence under minimal vector interactions.

We showed that nonminimal vector couplings have been used improperly in the
phenomenological description of elastic meson-nucleus scatterings potential
by observing that the four-current is not conserved when one uses either the
matrix $P\beta ^{\mu }$ or $\beta ^{\mu }P$, even though the linear forms
constructed from those matrices behave as true Lorentz vectors. We also
pointed out that the space component of the nonminimal vector potential can
not be absorbed into the spinor. Beyond that, we showed that the space
component of the nonminimal vector potential could be irrelevant for the
formation of bound states for potentials vanishing at infinity but its
presence is an essential ingredient for confinement.

For the one-dimensional problem, the DKP equation with nonminimal vector
potentials was mapped into a Sturm-Liouville problem in such a way that the
solution for linear potentials could be found by solving a Schr\"{o}%
dinger-like problem for the nonrelativistic harmonic oscillator. The
behavior of the solutions for this sort of DKP oscillator was discussed in
detail. That model reinforced the absence of Klein's paradox. Furthermore,
due to the fact that there is no room for the boson-antiboson production, a
boson embedded in this sort of background acquires an effective mass which
permits it can be strictly localized. We also showed that the DKP oscillator
for vector bosons is conditionally solvable.

In addition to provide a better understanding of the DKP theory with a sort
of coupling full of phenomenological relevance and not yet well explored in
the literature, it was conceived an exactly solvable vector model relating
to the confinement of bosons.

\bigskip

\bigskip\

\noindent \textbf{Acknowledgments}

This work was supported in part by means of funds provided by CAPES and
CNPq. The authors would like to thank an anonymous referee for drawing
attention to Ref. \cite{mr}, unavailable for us and partially cited in Ref.
\cite{gre}.

\newpage

\bigskip

\bigskip

\begin{thebibliography}{99}
\bibitem{pet} G. Petiau, Acad. R. Belg., A. Sci. M\'{e}m. Collect. \textbf{16%
}, No. 2 (1936); N. Kemmer, Proc. R. Soc. A \textbf{166}, 127 (1938); R. J.
Duffin, Phys. Rev. \textbf{54}, 1114 (1938).

\bibitem{kem} N. Kemmer, Proc. R. Soc. A \textbf{173}, 91 (1939).

\bibitem{now} M. Nowakowski, Phys. Lett. A \textbf{244}, 329 (1998).

\bibitem{lun} J. T. Lunardi \textit{et al.}, Phys. Lett. A \textbf{268}, 165
(2000).

\bibitem{mr} M. Riedel, \textit{Relativistische Gleichungen fuer
Spin-1-Teilchen}, Diplomarbeit, Institute for Theoretical Physics, Johann
Wolfgang Goethe-University, Frankfurt/Main (1979).

\bibitem{FIS} E. Fischbach, M. M. Nieto, and C. K. Scott, J. Math. Phys.
\textbf{14}, 1760 (1973).

\bibitem{fis1} E. Fischbach \textit{et al}., Phys. Rev. Lett. \textbf{26},
1200 (1971); E. Fischbach \textit{et al}., Phys. Rev. Lett. \textbf{27},
1407 (1971); E. Fischbach and M. M. Nieto, Phys. Rev. Lett. \textbf{29},
1046 (1972); N. G. Deshpande and P. C. McNamee, Phys. Rev. D \textbf{5},
1012 (1972); A. O. Barut and Z. Z. Aydin, Phys. Rev. D \textbf{6}, 3340
(1972); E. Fischbach, M. M. Nieto, and C. K. Scott, Phys. Rev. D \textbf{7},
207 (1973); Z. Z. Aydin and A. O. Barut, Phys. Rev. D \textbf{7}, 3522
(1973); M. D. Scadron and R. L. Thews, Phys. Rev. D \textbf{9}, 2180 (1974);
E. Fischbach \textit{et al}., Phys. Rev. D \textbf{9}, 2183 (1974); E.
Fischbach, M. M. Nieto, and C. K. Scott, Prog. Theor. Phys. \textbf{51},
1585 (1974); F. T. Meiere \textit{et al}., Phys. Rev. D \textbf{8}, 4209
(1973); E. Friedman, G. K\"{a}lbermann, and C. J. Batty, Phys. Rev. C
\textbf{34}, 2244 (1986).

\bibitem{cla1} B. C. Clark \textit{et al}., Phys. Rev. Lett. \textbf{55},
592 (1985).

\bibitem{kal} G. K\"{a}lbermann, Phys. Rev. C \textbf{34}, 2240 (1986); R.
E. Kozack \textit{et al}., Phys. Rev. C \textbf{37}, 2898 (1988); R. E.
Kozack, Phys. Rev. C \textbf{40}, 2181 (1989).

\bibitem{mis} V. K. Mishra \textit{et al}., Phys. Rev. C \textbf{43}, 801
(1991).

\bibitem{kur1} L. J. Kurth \textit{et al}., Phys. Rev. C \textbf{50}, 2624
(1994); R. C. Barret and Y. Nedjadi, Nucl. Phys. A \textbf{585}, 311c
(1995); L. J. Kurth \textit{et al}., Nucl. Phys. A \textbf{585}, 335c (1995).

\bibitem{ait} S. Ait-Tahar, J. S. Al-Khalili, and Y. Nedjadi, Nucl. Phys. A
\textbf{589}, 307 (1995).

\bibitem{cla2} B. C. Clark \textit{et al}., Phys. Lett. B \textbf{427}, 231
(1998).

\bibitem{gho1} P. Ghose, D. Home, and M. N. S. Roy, Phys. Lett. A \textbf{183%
}, 267 (1993).

\bibitem{gho2} P. Ghose, Phys. Lett. A \textbf{191}, 362 (1994); P. Ghose,
Found. Phys. \textbf{26}, 1441 (1996).

\bibitem{gho4} P. Ghose and M. K. Samal, Phys. Rev. E \textbf{64}, 036620
(2001).

\bibitem{gho5} P. Ghose\textit{\ et al.}, Phys. Lett. A \textbf{290}, 205
(2001).

\bibitem{bon} C. A. Bonin\textit{\ et al.}, quant-ph/0608002.

\bibitem{nik} H. Nikoli\'{c}, hep-th/0702060.

\bibitem{gho6} P. Ghose, M. K. Samal, and A. Datta, Phys. Lett. A \textbf{315%
}, 23 (2003).

\bibitem{stru} W. Struyve\textit{\ et al.}, Phys. Lett. A \textbf{322}, 84
(2004).

\bibitem{dat} A. Datta, \textit{High Spin Field Theories and Relativistic
Quantum Mechanics of Bosons}, in \textit{Bosons, Ferromagnetism and Crystal
Growth Research}, Horizons in World Physics, edited by E. Seifer (Nova
Publishers, New York, 2007) Vol. 257, Chap. 4, pp. 119-149.

\bibitem{tati} T. R. Cardoso, L. B. Castro, and A. S. de Castro, Phys. Lett.
A \textbf{372}, 5964 (2008).

\bibitem{gri} V. Gribov, Eur. Phys J. C \textbf{10}, 71 (1999).

\bibitem{kan} I. V. Kanatchikov, Rep. Math. Phys. \textbf{46}, 107 (2000).

\bibitem{fer} M. C. B. Fernandes and J. D. M. Viannna, Found. Phys. \textbf{%
29}, 201 (1999); A. O. Bolivar, Physica A \textbf{315}, 601 (2002).

\bibitem{red} V. M. Red%
\'{}%
kov, quant-ph/9812007; J. T. Lunardi, B. M. Pimentel, and R. G. Teixieira,
Gen. Rel. Grav. \textbf{34}, 491 (2002); R. Casana\textit{\ et al.}, Int. J.
Mod. Phys. A \textbf{17}, 4197 (2002); R. Casana\textit{\ et al.}, Gen. Rel.
Grav. \textbf{34}, 1941 (2002); R. Casana\textit{\ et al.}, Class. Quantum
Grav. \textbf{20}, 2457 (2003); R. Casana\textit{\ et al.}, Class. Quantum
Grav. \textbf{22}, 3083 (2003); K. Sogut and A. Havare, Class. Quantum Grav.
\textbf{23}, 7129 (2005); R. Casana, C. A. M. de Melo, and B. M. Pimentel,
Class. Quantum Grav. \textbf{24}, 723 (2007).

\bibitem{LUN} J. T. Lunardi\textit{\ et al.}, Int. J. Mod. Phys. A \textbf{17%
}, 205 (2002).

\bibitem{pim} B. M. Pimentel and V. Ya. Fainberg, Theor. Math. Phys. \textbf{%
124}, 1234 (2000); V. Ya. Fainberg and B. M. Pimentel, Phys. Lett. A \textbf{%
271}, 16 (2000).

\bibitem{mon} M. de Montigny\textit{\ et al.}, J. Phys. A \textbf{33}, L273
(2000); M. de Montigny\textit{\ et al.}, J. Phys. A \textbf{34}, 8901
(2001); M. C. B. Fernandes, A. E. Santana, and J. D. M. Viannna, J. Phys. A
\textbf{36}, 3841 (2003); J. D. M. Viannna, M. C. B. Fernandes, and A. E.
Santana, Found. Phys. \textbf{35}, 109 (2005); E. S. Santos and L. M. Abreu,
J. Phys. A \textbf{41}, 075407 (2008).

\bibitem{CAS1} R. Casana \textit{et al.}, hep-th/0506193.

\bibitem{CAS2} R. Casana \textit{et al.}, Phys. Lett. A \textbf{316}, 33
(2003).

\bibitem{dai} J. Daicic and N. E. Frankel, J. Phys. A \textbf{26}, 1397
(1993); K. Sogut, A. Havare, and I. Acikgoz, J. Math. Phys. \textbf{43},
3952 (2002).

\bibitem{swa} J. A. Swansson and B. H. J. McKellar, J. Phys. A \textbf{34},
1051 (2001).

\bibitem{BOU1} A. Boumali, Can. J. Phys. \textbf{82}, 67 (2004); A. Boumali,
Can. J. Phys. \textbf{85}, 1417 (2007).

\bibitem{mer2} M. Merad, Int. J. Theor. Phys. \textbf{46}, 2105 (2007); T.
R. Cardoso, L. B. Castro, and A. S. de Castro, hep-th/0908.0016 (to appear
in Int. J. Theor. Phys.).

\bibitem{mer1} M. Merad, H. Bada, and A. Lecheheb, Czech. J. Phys. \textbf{56%
}, 765 (2006).

\bibitem{deb} N. Debergh, J. Ndimubandi, and D. Strivay, Z. Phys. C \textbf{%
56}, 421 (1992); Y. Nedjadi and R. C. Barret, J. Phys. A \textbf{27}, 4301
(1994).

\bibitem{Ait} Y. Nedjadi, S. Ait-Tahar, and R. C. Barret, J. Phys. A \textbf{%
31}, 3867 (1998).

\bibitem{NB} Y. Nedjadi and R. C. Barret, J. Phys. A \textbf{31}, 6717
(1998).

\bibitem{kuli} D. A. Kulikov, R. S. Tutik, and A. P. Yaroshenko, Mod. Phys.
Lett. A \textbf{20}, 43 (2005).

\bibitem{boucheto} A. Boumali and L. Chetouani, Phys. Lett. A \textbf{346},
261 (2005).

\bibitem{BOZ10} I. Boztosun \textit{\ et al.}, J. Math. Phys. \textbf{47},
062301 (2006).

\bibitem{bou3} A. Boumali, J. Math. Phys. \textbf{49}, 022302 (2008).

\bibitem{kas} Y. Kasri and L. Chetouani, Int. J. Theor. Phys. \textbf{47},
2249 (2008).

\bibitem{BOU4} A. Boumali, Phys. Scr. \textbf{76}, 669 (2007).

\bibitem{yas1} F. Ya\c{s}uk\textit{,} M. Karakoc, and I. Boztosun, Phys.
Scr. \textbf{78}, 045010 (2008).

\bibitem{CHE} L. Chetouani\textit{\ et al.}, Int. J. Theor. Phys. \textbf{43}%
, 1147 (2004).

\bibitem{bou} \noindent B. Boutabia-Ch\'{e}raitia and T. Boudjedaa, Phys.
Lett. A \textbf{338}, 97 (2005).

\bibitem{yas} F. Ya\c{s}uk\textit{\ et al.}, Phys. Scr. \textbf{71}, 340
(2005).

\bibitem{ned2} Y. Nedjadi and R. C. Barret, J. Math. Phys. \textbf{35}, 4517
(1994).

\bibitem{ned1} Y. Nedjadi and R. C. Barret, J. Phys. G \textbf{19}, 87
(1993).

\bibitem{gon} S. G\"{o}nen, A. Havare, and N. Unal, hep-th/0207087.

\bibitem{ccc2} T. R. Cardoso, L. B. Castro, and A. S. de Castro,
hep-th/0905.0427 (to appear in Can. J. Phys.); see also T. R. Cardoso, L. B.
Castro, and A. S. de Castro, Can. J. Phys. \textbf{87}, 857 (2009).

\bibitem{puk} H. M. Pilkuhn, \textit{Relativistic Quantum Mechanics}
(Springer, Berlin, 2003) 2nd ed.

\bibitem{gue} See, \textit{e.g.}, R. F. Guertin and T. L. Wilson, Phys. Rev.
D \textbf{15}, 1518 (1977).

\bibitem{vij} See, \textit{e.g.}, B. Vijayalakshmi, M. Seetharaman, and P.
M. Mathews, J. Phys. A \textbf{12}, 665 (1979).

\bibitem{sch} L. I. Schiff, \textit{Quantum Mechanics }(McGraw-Hill, New
York, 1968) 3rd ed.

\bibitem{gre} W. Greiner, \textit{Relativistic Quantum Mechanics: Wave
Equations} (Springer, Berlin, 1990).

\bibitem{str} P. Strange, \textit{Relativistic Quantum Mechanics with
Applications in Condensed Matter and Atomic Physics} (Cambridge University
Press, Cambridge, 1998).

\bibitem{mos} A. Mostafazadeh and F. Zamani, quant-ph/0312078; A.
Mostafazadeh and F. Zamani, Ann. Phys. (N.Y.) \textbf{321}, 2183 (2006); A.
Mostafazadeh and F. Zamani, Ann. Phys. (N.Y.) \textbf{321}, 2210 (2006); F.
Kleefeld, Czech. J. Phys. \textbf{56}, 999 (2006).
\end{thebibliography}
\end{document}